\documentclass[conference]{IEEEtran}

\usepackage[utf8]{inputenc}
\usepackage{fontenc}
\usepackage{amsmath, amssymb, amsfonts}
\usepackage{hyperref}
\usepackage{booktabs}
\usepackage{multirow}
\usepackage{url}
\usepackage{caption}
\usepackage{graphicx}

\usepackage{dblfloatfix}   
\usepackage{microtype}     

\usepackage{tabularx}
\usepackage{adjustbox}
\usepackage{threeparttable}
\usepackage{array}
\newcolumntype{Y}{>{\raggedright\arraybackslash}X}
\usepackage{balance}

\title{\textbf{Bilevel Optimization for Covert Memory Tampering in Heterogeneous Multi-Agent Architectures (XAMT)}}
\author{%
  Akhil Sharma\thanks{Chief Scientist, SnowCrashLabs, San Francisco, USA}%
  \and
  Shaikh Yaser Arafat\thanks{Toronto, Canada}%
  \and
  Jai Kumar Sharma\thanks{New Delhi, India}%
  \and
  Ken Huang\thanks{Chief AI Officer, Distributed Apps, Washington DC, USA}%
}
\date{}

\begin{document}
\maketitle

\begin{abstract}
The increasing operational reliance on complex Multi-Agent Systems (MAS) across safety-critical domains necessitates a rigorous assessment of their adversarial robustness. Modern MAS are inherently heterogeneous, often integrating conventional machine learning paradigms, such as Multi-Agent Reinforcement Learning (MARL), with emerging Large Language Model (LLM) agent architectures utilizing Retrieval-Augmented Generation (RAG). A critical, shared vulnerability across these disparate systems is their reliance on centralized, externalized memory components, whether the shared Experience Replay (ER) buffer in MARL or the external Knowledge Base ($\mathcal{K}$) in RAG agents \cite{Ref1, Ref2}. This paper proposes the \textbf{XAMT (Bilevel Optimization for Covert Memory Tampering in Heterogeneous Multi-Agent Architectures)} framework. XAMT formally models the attack generation as a novel bilevel optimization (BO) problem. The Upper Level minimizes the magnitude of the perturbation ($\delta$), effectively enforcing covertness, while simultaneously maximizing the resulting divergence of the system's behavior (Lower Level) toward an adversary-defined target policy or response. We provide rigorous mathematical instantiations of XAMT for two structurally distinct targets: Centralized Training Decentralized Execution (CTDE) MARL algorithms and RAG-based LLM agents. Our theoretical analysis confirms that BO is uniquely suited for crafting stealthy, minimal-perturbation poisons that circumvent existing detection heuristics. We propose comprehensive experimental protocols utilizing established security benchmarks, including the StarCraft Multi-Agent Challenge (SMAC) and SafeRAG, to quantitatively demonstrate XAMT's high effectiveness and sub-percent poison rates (e.g., $\rho \le 1\%$ in MARL and $\rho \le 0.1\%$ in RAG regimes). This research defines a new, unified class of training-time threats essential for developing intrinsically secure MAS.
\end{abstract}


\section{Introduction}

\begin{figure*}[!t]
    \centering
    \includegraphics[width=\textwidth,keepaspectratio]{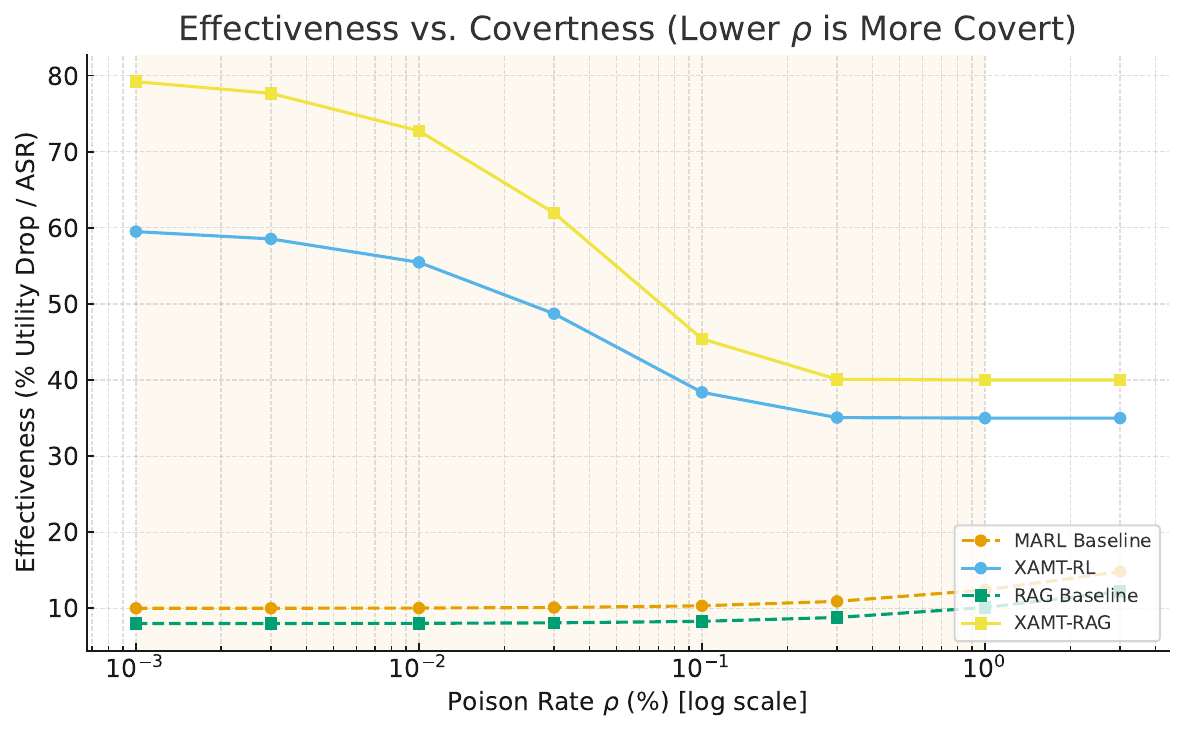}
    \caption{\textbf{XAMT at a glance—effectiveness vs. covertness.}
    \emph{Left:} Attack Success / Utility Drop versus poison rate ($\rho$, log scale) for MARL (XAMT-RL) and RAG (XAMT-RAG) compared to non-optimized baselines. XAMT achieves high impact at sub-percent $\rho$.
    \emph{Right:} Semantic deviation (proxy for detectability in RAG) versus ASR. XAMT-RAG attains higher ASR with lower semantic drift, reflecting the bilevel “minimal perturbation” objective.}
    \label{fig:intro_tradeoffs}
\end{figure*}

\subsection{The Ubiquity and Fragility of Heterogeneous Multi-Agent Systems}
Multi-Agent Systems (MAS) have transitioned from theoretical constructs to integral components in real-world, high-stakes environments. These applications range widely, encompassing smart transportation networks, autonomous vehicle coordination, advanced cyber defense mechanisms, and complex economic modeling \cite{Ref3, Ref4, Ref5}. The deployment of these systems in critical infrastructure mandates that their decision-making processes be robust, reliable, and trustworthy.

A notable architectural shift is the rising heterogeneity within MAS. Contemporary systems rarely rely on a single, monolithic learning paradigm. Instead, they often integrate traditional learning components, such as cooperative MARL agents utilizing shared state representations or centralized critics \cite{Ref6}, with cutting-edge components, such as LLM agents that enhance their reasoning and contextual awareness via RAG processes \cite{Ref2, Ref7}. While this integration leverages diverse capabilities to tackle complex, layered tasks \cite{Ref8}, it concurrently introduces complex interdependencies and expanded vulnerability surfaces.

This architectural complexity demands sophisticated mechanisms for coordination and knowledge sharing. Successful cooperation in MAS is predicated on the ability of individual agents to align their actions, often through accessing consistent, shared information \cite{Ref9}. This critical dependency highlights a shared weakness: the components responsible for storing and mediating this shared knowledge.

\subsection{Memory as the Unified Critical Attack Surface}
The integrity of memory is paramount to collective intelligence in MAS. This shared information infrastructure acts as the system's "computational exocortex"—a dynamic, systematic process that stores, retrieves, and synthesizes experiences crucial for coordination \cite{Ref1}. Disturbingly, this collective memory layer represents a unified, high-leverage attack surface across heterogeneous architectures.

In the MARL domain, cooperative success hinges on frameworks like Centralized Training, Decentralized Execution (CTDE) \cite{Ref6, Ref10}. These methods typically rely on a centralized critic function, which is trained using a shared Experience Replay (ER) buffer ($\mathcal{D}$) containing collective state-action histories \cite{Ref11}. Corrupting the transitions stored in this buffer, particularly through reward or state poisoning, directly compromises system integrity and resilience \cite{Ref12, Ref13}. The centralized critic, while improving stability and credit assignment, becomes a single point of failure where a localized tamper can affect the global policy update \cite{Ref6}.

Similarly, in LLM-driven agents, the external RAG knowledge base ($\mathcal{K}$) functions as persistent, external memory \cite{Ref2}. This external data store, often sourced from public or proprietary documents, presents a novel attack vector distinct from the challenges of core model weight manipulation \cite{Ref14, Ref15}. Attacks here involve injecting malicious, low-quality, or misleading texts into $\mathcal{K}$ \cite{Ref7, Ref16}. Systemic security failure, therefore, often results not from the compromise of a single agent's reasoning capabilities, but from the corruption of the foundational shared context or experience upon which the entire collective relies \cite{Ref1}.

\subsection{The Need for Covert, Optimization-Driven Attacks}
Traditional poisoning attacks, while often effective, frequently involve large-scale manipulations, such as extensive label flipping, high-magnitude reward alterations, or the introduction of a high percentage of synthetic samples \cite{Ref12, Ref17}. Such attacks are inherently detectable via anomaly detection systems or simple threshold checks designed to detect massive distributional shifts \cite{Ref18}. For an attack to pose a truly sophisticated threat in a monitored, production environment, it must prioritize \textit{covertness}. A covert attack necessitates that the perturbation $\delta$ introduced into the memory system must be minimal (e.g., small $L_p$ norm on numerical data) or semantically plausible (clean-label text) \cite{Ref19, Ref20}.

Achieving maximal functional damage with minimal detectable perturbation is a computationally challenging constraint satisfaction problem. Bilevel Optimization (BO) provides the necessary mathematical rigor to solve this highly constrained, hierarchical adversarial problem \cite{Ref21, Ref22}. BO allows the attacker to explicitly model the victim system's learning process (the Lower Level) as a constraint on the overall attack objective (the Upper Level) \cite{Ref23}. This capability is critical because it permits the selection of the \textit{minimum required perturbation} $\delta$ that guarantees the maximal target utility drop after the victim completes its training iteration $\theta^*(\delta)$ \cite{Ref24, Ref25}. BO is proven effective in generating clean-label poisoning points with imperceptible distortions to reduce certified robustness guarantees, establishing it as the ideal mathematical tool for fulfilling the \textit{covert} requirement of the XAMT framework \cite{Ref19, Ref20}.

\subsection{Contributions and Novelty}
The XAMT framework makes the following critical contributions to adversarial machine learning and MAS security research:

\begin{enumerate}
    \item \textbf{Unified Framework:} The paper proposes XAMT, which is, to our knowledge, the first unified bilevel optimization framework that explicitly spans covert poisoning across structurally heterogeneous memory components ($\mathcal{M}$) in MAS, specifically CTDE-style MARL experience replay buffers and RAG knowledge bases.
    \item \textbf{Formalization of Covertness:} A generalized definition of the Minimal Perturbation Constraint $R(\delta)$ is formalized, tailored to the modality of $\mathcal{M}$. This constraint uses $L_p$ norms for numerical MARL data and semantic distance ($D_{\text{sem}}$) metrics for textual RAG data, rigorously integrating stealth into the BO objective.
    \item \textbf{Dual Instantiations:} Detailed mathematical specifications are provided for XAMT-RL (targeting centralized critics in CTDE) and XAMT-RAG (targeting knowledge bases), demonstrating the framework's broad applicability.
    \item \textbf{Experimental Protocol:} Comprehensive, comparative experimental protocols are detailed using established security benchmarks, including SMAC and SafeRAG, to quantify the effectiveness and covertness of XAMT against state-of-the-art multi-agent and RAG architectures.
\end{enumerate}

\section{Formal Threat Model and Assumptions}
\label{sec:threat_model}

We formalize the adversarial scenario underpinning XAMT through a threat model that defines attacker capabilities, victim system properties, and the operational scope of memory tampering attacks. This model provides the necessary context for the bilevel optimization framework and clarifies the practical feasibility of the proposed threats.

\subsection{Attacker Capabilities}
\textbf{Attacker notation:} we denote the attacker as $\mathcal{A}$.

The attacker is modeled as a \textbf{capable but constrained adversary} with the following explicit capabilities.

\subsubsection{Knowledge Model}
$\mathcal{A}$ operates under a \textbf{white-box} assumption regarding the victim's learning architecture but not necessarily its runtime data streams. Specifically:

\begin{itemize}
\item \textbf{Structural Knowledge:} $\mathcal{A}$ knows the victim's algorithmic class (e.g., QMIX/MAPPO for MARL; dense retrieval with a specific LLM for RAG), hyperparameters (including $\lambda$), learning-rate schedules, and the centralized memory architecture $\mathcal{M}$ \cite{Ref12, Ref14}.
\item \textbf{Training Data Distribution:} $\mathcal{A}$ has access to a surrogate dataset $\mathcal{D}_{\text{surrogate}}$ that approximates the distribution of the victim's clean memory $\mathcal{M}$. This is a standard assumption in data poisoning research, enabling the attacker to simulate the lower-level optimization \cite{Ref18, Ref19}.
\item \textbf{No Real-time Observability:} $\mathcal{A}$ does \textit{not} require real-time observation of the victim's training trajectory or access to intermediate gradients during execution. This distinguishes XAMT from dynamic adversarial examples and focuses on \textbf{training-time poisoning}.
\end{itemize}

\subsubsection{Access Model}
$\mathcal{A}$ possesses \textbf{write-access} to the centralized memory component $\mathcal{M}$ with the following constraints:

\begin{itemize}
\item \textbf{Injection Point:} $\mathcal{A}$ can inject perturbations $\delta$ into $\mathcal{M}$ \textit{prior to or during} the victim's training phase, but cannot modify the victim's model parameters $\theta$ directly \cite{Ref13, Ref25}. For MARL, this corresponds to tampering with offline-collected experience batches before they enter the replay buffer. For RAG, this corresponds to injecting documents into $\mathcal{K}$ before retriever fine-tuning or before inference-time retrieval.
\item \textbf{Rate Constraint:} The injection quantity is limited to a sub-percent fraction:
\[
\rho = \frac{|\delta|}{|\mathcal{M}|} \le 1\% \;\; \text{(MARL)}, \qquad \rho \le 0.1\% \;\; \text{(RAG)}
\]
\cite{Ref7, Ref13}. This reflects realistic scenarios where large-scale corruption would trigger integrity checks.
\item \textbf{Covertness Mandate:} $\mathcal{A}$ must ensure $\delta$ evades \textit{static, heuristic detection}: anomaly thresholds on $L_p$ norms for numerical data \cite{Ref24} and semantic filters (e.g., perplexity or embedding proximity checks) for textual data \cite{Ref29, Ref37}.
\end{itemize}

\subsubsection{Computational Resources}
$\mathcal{A}$ has sufficient compute to solve the bilevel problem:

\begin{itemize}
\item \textbf{Hessian Approximation:} Access to GPU-accelerated solvers (e.g., PBGD, BIGRAD) capable of approximating $\nabla_{\delta} L_{\mathcal{A}}$ via implicit differentiation \cite{Ref21, Ref23}.
\item \textbf{Surrogate Training:} Ability to run multiple inner-loop training episodes on $\mathcal{D}_{\text{surrogate}}$ to estimate $\theta^*(\delta)$ during the outer-loop optimization \cite{Ref20, Ref25}.
\end{itemize}

\subsection{Victim System Assumptions}
\textbf{Victim notation:} we denote the victim system as $\mathcal{S}$.

The target system is assumed to be a \textbf{benign, well-intentioned learner} with these properties.

\subsubsection{Learning Process Integrity}
The victim executes its standard training algorithm $\mathcal{V}$ without adversarial hardening:

\begin{itemize}
\item \textbf{Optimization Fidelity:} $\mathcal{V}$ faithfully minimizes $L_{\mathcal{S}}$ via stochastic gradient descent (e.g., Adam, RMSProp) on the poisoned memory $\mathcal{M}+\delta$ \cite{Ref12, Ref42}.
\item \textbf{No Poison-Aware Defenses:} $\mathcal{S}$ does \textit{not} employ Byzantine-robust aggregation, data sanitization, or adversarial training during the initial training phase \cite{Ref18, Ref30}. This reflects the status quo in most open-source MARL/RAG implementations.
\end{itemize}

\subsubsection{Memory Architecture Vulnerabilities}
$\mathcal{S}$ relies on \textbf{centralized, weakly protected memory}:

\begin{itemize}
\item \textbf{CTDE MARL:} Uses a shared experience replay buffer $\mathcal{D}$ accessible to centralized critic updates. No per-agent isolation or integrity checks on stored transitions \cite{Ref6, Ref11}.
\item \textbf{RAG Systems:} Employs a flat knowledge base $\mathcal{K}$ with vector-based retrieval. No robust provenance verification or retrieval-time safety filtering beyond semantic similarity thresholds \cite{Ref2, Ref14}.
\end{itemize}

\subsubsection{Detection Mechanisms}
$\mathcal{S}$ employs only \textbf{superficial, post-hoc integrity checks}:

\begin{itemize}
\item \textbf{Numerical Monitoring:} Simple threshold-based anomaly detection on reward/state magnitudes (e.g., $ \|\delta_{RL}\|_{\infty} < \epsilon_{\text{detect}}$) \cite{Ref24}.
\item \textbf{Textual Filtering:} Basic semantic filters that flag documents with high perplexity or low BERTScore relative to the corpus \cite{Ref29, Ref37}. XAMT explicitly optimizes $R(\delta)$ to remain below such heuristics.
\end{itemize}

\subsection{System Scope and Limitations}
The XAMT threat model is bounded by the following assumptions to ensure practical relevance.

\subsubsection{Temporal Scope}
Attacks are \textbf{training-time only}. Evasion of runtime adversarial detection (e.g., during MARL execution or RAG inference) is out-of-scope, focusing on poisoning foundational memory before deployment \cite{Ref13, Ref25}.

\subsubsection{Architectural Scope}
\begin{itemize}
\item \textbf{MARL:} Limited to CTDE algorithms with centralized critics (e.g., QMIX, MAPPO, VDN). Fully decentralized or purely on-policy methods without shared buffers (e.g., independent Q-learning) are not direct targets \cite{Ref6, Ref10}.
\item \textbf{RAG:} Focuses on dense retrieval with vector embeddings. Sparse retrieval (e.g., BM25) or graph-based retrieval requires alternative definitions of $D_{\text{sem}}$, but remains compatible with the BO framework \cite{Ref42, Ref48}.
\end{itemize}

\subsubsection{Attack Goal Scope}
$\mathcal{A}$ optimizes for a \textbf{single, predefined target}:

\begin{itemize}
\item \textbf{MARL:} A target policy $T$ that induces measurable utility drop (e.g., win-rate decline) \cite{Ref36, Ref41}.
\item \textbf{RAG:} A specific adversarial response $Y_T$ to a trigger prompt $P_{tr}$, measured via ASR \cite{Ref7, Ref37}.
\end{itemize}

Multi-target or dynamic objectives are deferred to future work.

\subsection{Success Metrics and Attack Objectives}

\begin{table}[!t]
\centering
\caption{Threat Model Success Criteria}
\label{tab:threat_model_metrics}
\footnotesize
\setlength{\tabcolsep}{3pt} 
\renewcommand{\arraystretch}{1.15} 

\begin{adjustbox}{max width=\columnwidth}
\begin{tabularx}{\columnwidth}{|p{0.22\columnwidth}|X|X|}
\hline
\textbf{Metric} & \textbf{XAMT-RL (MARL)} & \textbf{XAMT-RAG (LLM)} \\
\hline
\textbf{Effectiveness} &
Utility Drop $\ge 40\%$ at $\rho \le 1\%$ &
ASR $\ge 90\%$ at $\rho \le 0.1\%$ \\
\hline
\textbf{Covertness} &
$\|\delta_{RL}\|_{\infty} < 0.05$, $\|\delta_{RL}\|_2 < 0.1$ &
$D_{\text{sem}} < 0.15$, Perplexity $\le$ Baseline $+10\%$ \\
\hline
\textbf{Evasion} &
Bypass threshold $\epsilon_{\text{detect}} = 0.1$ &
Evade BERTScore filter $\tau = 0.85$ \\
\hline
\end{tabularx}
\end{adjustbox}
\end{table}

$\mathcal{A}$ succeeds iff: (1) the effectiveness threshold is met, (2) the covertness constraints are satisfied, and (3) the perturbation evades victim detection heuristics. This tripartite criterion is \textit{explicitly encoded} in the XAMT bilevel objective (Section~\ref{sec:xamt_framework}), where $\lambda$ balances $L_{\mathcal{A}}$ and $R(\delta)$ to satisfy all constraints simultaneously.

\section{Background and Related Work}

\subsection{Foundations of Bilevel Optimization in Adversarial Machine Learning}
Bilevel optimization (BO) is a hierarchical optimization paradigm characterized by two nested optimization problems. The solution to the inner (lower-level) problem constrains or dictates the objective of the outer (upper-level) problem \cite{Ref21}. This structure is essential for modeling adversarial interactions in machine learning, particularly where the attacker's actions influence the target model's training dynamics \cite{Ref23}.

In the context of data poisoning, the victim model's training process is formalized as the lower-level problem, which produces the optimal model parameters $\theta^*(\delta)$ given the poisoned data $\mathcal{M}+\delta$ \cite{Ref18}. The attacker (Upper Level) then optimizes the poison $\delta$ to maximize the adversarial loss $L_{\mathcal{A}}(\theta^*)$, subject to the constraint of minimal perturbation, $\lambda R(\delta)$ \cite{Ref24}. Previous research has effectively leveraged BO to generate clean-label poisoning points with minimal input edits, achieving high targeted damage—such as reducing the average certified radius (ACR) of a target class by over $30\%$—without massive accuracy reductions that would alert the victim \cite{Ref19, Ref20}. This body of work confirms that BO is not merely a tool for optimization but a necessary methodology for achieving the required stealth and precision mandated by the XAMT framework. Furthermore, integrating BO within deep learning frameworks has been made feasible through recent advances like Differentiating through Bilevel Optimization Programming (BIGRAD), which extends existing single-level optimization programming approaches to handle the complex gradient calculations required by nested problems \cite{Ref23}.

\subsection{Adversarial Attacks in Multi-Agent Reinforcement Learning (MARL)}
Multi-Agent Reinforcement Learning (MARL) models are increasingly recognized as vulnerable to adversarial perturbations \cite{Ref26}. Attacks often manifest through manipulating agent states, actions, or rewards during training or execution \cite{Ref3, Ref27}.

A significant subset of MARL algorithms adheres to the Centralized Training, Decentralized Execution (CTDE) paradigm \cite{Ref6, Ref10}. This approach, while facilitating improved coordination by granting the critic access to global information (often including the true system state) during training, creates centralized components that are single points of failure \cite{Ref6}. The centralized critic's reliance on the shared Experience Replay (ER) buffer $\mathcal{D}$ makes this memory unit a critical attack target \cite{Ref11}.

Reward poisoning in offline MARL environments has been shown to be particularly effective, where modifying the rewards in a pre-collected dataset can compel rational agents to adopt a malicious target policy, often resulting in a Markov Perfect Equilibrium (MPE) under the poisoned dynamics \cite{Ref12, Ref28}. The architectural selection of CTDE, specifically the dependence on centralized critics, inherently creates a high-leverage target within the shared experience buffer. A localized, covert memory tamper $\delta_{RL}$ within this buffer can disproportionately corrupt the global policy update, making it an efficient mechanism for policy manipulation. This exploitation is further observed in attacks that are stealthy during offline training but trigger catastrophic failures during subsequent online fine-tuning (O2O) \cite{Ref13, Ref25}. These sophisticated attacks often leverage BO techniques to promote value over-estimation or distribution shifts within the critic network, effectively exploiting the centralized critic's potential for bias and variance \cite{Ref6, Ref25}. Existing MARL poisoning work, however, typically focuses on reward or state manipulation within MARL alone and does not provide a cross-domain, unified treatment that also regularizes covertness across heterogeneous memory modalities as a first-class objective.

\subsection{Adversarial Attacks in LLM Agent Memory (RAG)}
With the rise of agentic systems driven by Large Language Models (LLMs), new vulnerabilities related to external memory have emerged. Retrieval-Augmented Generation (RAG) systems, designed to ground LLM responses in external knowledge, introduce the knowledge base ($\mathcal{K}$) as a primary attack surface \cite{Ref14, Ref15}. Attacks such as AgentPoison and PoisonedRAG inject malicious demonstrations or texts directly into the RAG system's long-term memory or knowledge base \cite{Ref7, Ref14}.

For RAG attacks to be effective and evade sophisticated detection systems (e.g., retrieval safety assessments such as RevPRAG \cite{Ref2, Ref37}), the injected text must be highly covert \cite{Ref29}. This requires maintaining the semantic and syntactic integrity of the poisoned text while maximizing the probability of retrieval given a specific trigger \cite{Ref30}. Covert textual poisoning relies on techniques such as clean-label attacks and synonym substitution, where only minimal text insertions or replacements are used to maintain fluency and plausibility \cite{Ref31, Ref32}. This architectural shift from numerical, vector-based optimization (as in MARL states) to linguistic/semantic optimization (as in RAG text) means the definition of covertness must evolve from a simple $L_p$ norm constraint to a metric based on semantic distance ($D_{\text{sem}}$) \cite{Ref29}. Prior work such as AgentPoison and PoisonedRAG focuses exclusively on the RAG setting (e.g., constrained trigger and document optimization) and does not offer a unified bilevel formulation that simultaneously spans CTDE-MARL ER buffers and RAG KBs with a shared, modality-aware covertness regularization. XAMT is designed to fill exactly this cross-domain gap.

\subsection{The Convergence Gap: Justification for XAMT}
Existing adversarial machine learning research tends to segregate threats into distinct domains, focusing either on MARL policy manipulation or LLM/RAG knowledge corruption. A critical research gap exists in providing a unified, optimization-based framework that treats the memory unit $\mathcal{M}$ generically, while simultaneously addressing the specific dependencies and modalities of heterogeneous agent architectures.

The XAMT framework is designed to bridge this gap. By leveraging Bilevel Optimization, XAMT provides a mathematically robust method to guarantee that minimal perturbations in $\mathcal{M}$ (regardless of whether $\mathcal{M}$ is a numerical vector or textual document) achieve the maximal targeted strategic damage across diverse multi-agent learning paradigms. This unified approach is necessary for designing defenses capable of addressing systemic vulnerabilities across modern, complex MAS deployments.

\section{The XAMT Framework: Formalizing Covert Memory Tampering}

The XAMT framework mathematically formalizes the adversarial interaction between a stealthy attacker and a target MAS learning process utilizing shared memory. The attack is defined by the objective of maximizing a predefined security failure metric (effectiveness) while strictly adhering to a constraint on the perturbation magnitude (covertness).

\subsection{System Model and Attacker Goals}
The target system is modeled as a \textbf{Heterogeneous Multi-Agent System ($\mathcal{S}$)} comprising a collection of agents, $\mathcal{A} = \{A_1, \dots, A_N\}$, interacting within an environment $\mathcal{E}$. All agents rely on an \textbf{Abstract System Memory ($\mathcal{M}$)} for learning and coordination. $\mathcal{M}$ serves as the generic term for the critical, centralized information repository: either the centralized experience buffer $\mathcal{D}$ (for MARL) or the retrieval knowledge base $\mathcal{K}$ (for RAG).

The system updates its internal parameters $\theta$ (which may represent policies $\pi$ in MARL or LLM/retriever weights in RAG) using a \textbf{Victim Learning Process ($\mathcal{V}$)}, such that the resulting optimal parameters are $\theta^* = \mathcal{V}(\mathcal{M})$.

The objective of the \textbf{Covert Tampering Attacker ($\mathcal{A}$)} is to introduce a minimal perturbation $\delta$ into $\mathcal{M}$. This perturbation must be optimized such that the resulting system parameters $\theta^*(\mathcal{M}+\delta)$ maximally deviate toward a malicious target behavior $T$.

\subsection{The XAMT Bilevel Optimization Formulation}
The attacker's strategy is inherently hierarchical, leading to the generalized bilevel optimization (BO) formulation for memory perturbation:
\begin{equation}
\min_{\delta}\; L_{\mathcal{A}}(\theta^*(\delta)) + \lambda R(\delta)
\label{eq:xamt_upper}
\end{equation}
subject to
\begin{equation}
\theta^*(\delta) \in \arg\min_{\theta} L_{\mathcal{S}}(\theta, \mathcal{M} + \delta).
\label{eq:xamt_lower}
\end{equation}

Equation~\eqref{eq:xamt_upper} defines the \emph{upper-level attacker objective}, while
Equation~\eqref{eq:xamt_lower} represents the \emph{lower-level system learning process}.

\begin{enumerate}
    \item \textbf{Lower-Level Objective ($L_{\mathcal{S}}$):} This function represents the benign training or updating process of the victim system. In MARL, $L_{\mathcal{S}}$ might be the policy iteration loss (e.g., QMIX or MAPPO loss) \cite{Ref6}. In RAG, $L_{\mathcal{S}}$ could model the training loss for the retriever embeddings or the maximum likelihood loss of the generator. The solution $\theta^*(\delta)$ represents the system parameters optimized on the corrupted memory $\mathcal{M} + \delta$.
    \item \textbf{Upper-Level Objective ($L_{\mathcal{A}}$):} This loss quantifies the attacker's success, aiming to maximize the divergence of the resulting optimal parameters $\theta^*(\delta)$ from the desired system performance. For MARL, this involves maximizing the negative utility (e.g., minimizing win rate \cite{Ref36}). For RAG, this means maximizing the probability of generating a specific target response \cite{Ref37}.
    \item \textbf{Covertness Constraint ($R(\delta)$):} This is the crucial regularization term, governed by the hyperparameter $\lambda > 0$, which explicitly balances the trade-off between attack effectiveness ($L_{\mathcal{A}}$) and the cost of manipulation \cite{Ref24}. $R(\delta)$ is specialized based on the memory modality:
  \begin{equation}
R(\delta) =
\begin{cases}
\|\delta\|_p^p, & \mathcal{M}\ \text{numerical} \\
D_{\text{sem}}(\mathcal{M}, \mathcal{M}+\delta), & \mathcal{M}\ \text{textual}
\end{cases}
\label{eq:covertness}
\end{equation}

where numerical memory corresponds to MARL experience buffers, and textual memory corresponds to RAG knowledge bases.

    For numerical data (e.g., states or rewards), $L_p$ norms (typically $L_\infty$ or $L_2$) ensure that the perturbation magnitude remains below detection thresholds \cite{Ref24}. For textual data, the semantic distance $D_{\text{sem}}$ (calculated via embedding differences or linguistic models) guarantees the malicious text maintains semantic plausibility, satisfying the clean-label requirement \cite{Ref29}.
\end{enumerate}

\subsection{Computational Solution and Differentiability}
The primary computational challenge in solving the XAMT BO formulation lies in computing the gradient of the upper-level objective with respect to the perturbation $\delta$, specifically $\nabla_{\delta} L_{\mathcal{A}}$. Since the lower-level solution $\theta^*(\delta)$ is implicitly defined as the result of a potentially non-convex optimization process, traditional gradient methods are insufficient. The Implicit Function Theorem (IFT) approach is required \cite{Ref38}.

This differentiation necessitates calculating the gradient flow from the lower-level solution $\theta^*(\delta)$ back to the upper-level variable $\delta$ \cite{Ref38}. This typically involves the inverse Hessian ($H^{-1}$) of the lower-level loss function $L_{\mathcal{S}}$ with respect to $\theta$:
$$
\nabla_{\delta} L_{\mathcal{A}}(\theta^*(\delta)) = - \nabla_{\delta, \theta} L_{\mathcal{S}}(\theta^*(\delta), \delta)^T H^{-1} \nabla_{\theta} L_{\mathcal{A}}(\theta^*(\delta))
$$
In large-scale deep learning applications prevalent in MARL and LLM systems, the direct computation or inversion of the Hessian matrix $H$ can be computationally intractable \cite{Ref39}. The resulting memory and time complexity limits the scalability of minimal-perturbation attacks \cite{Ref39}.

To address these limitations and render XAMT practical for large-scale, non-convex problems, specialized optimization techniques are adopted. Methods such as Penalty-Based Bilevel Gradient Descent (PBGD) transform the constrained bilevel problem into a sequence of more manageable single-level optimization problems (often minimax problems), improving scalability without requiring strict lower-level strong convexity \cite{Ref21, Ref22}. Furthermore, differentiating through bilevel problems can utilize techniques like BIGRAD, which leverages vector-Jacobian products for efficient gradient calculation, enabling end-to-end learning within modern machine learning frameworks \cite{Ref23}. The practicality of XAMT is predicated on the ability of these specialized solvers to efficiently approximate the inverse Hessian, extending the applicability of BO to non-convex, high-dimensional parameter spaces common in deep MARL and RAG systems.

Table \ref{tab:xamt_unified} summarizes the generalized components of the XAMT framework.

\begin{table*}[!t]
\centering
\caption{Unified XAMT Framework Definition}
\label{tab:xamt_unified}
\begin{threeparttable}
\small
\begin{adjustbox}{max width=\textwidth}
\begin{tabularx}{\textwidth}{|Y|Y|Y|Y|}
\toprule
\textbf{Component} & \textbf{Upper-Level (Attacker $\mathcal{A}$ Loss)} & \textbf{Lower-Level (Victim System $\mathcal{S}$ Loss)} & \textbf{Covertness ($R(\delta)$)} \\
\midrule
\textbf{Objective} & $\min_{\delta} L_{\mathcal{A}}(\theta^*(\delta)) + \lambda R(\delta)$ & $\theta^*(\delta) \in \text{argmin}_{\theta} L_{\mathcal{S}}(\theta, \mathcal{M} + \delta)$ & Minimization of Perturbation Cost $R(\delta)$ \\
\textbf{Goal} & Drive system to target policy/response ($T$) with minimal input change ($\delta$). & Find optimal parameters $\theta$ that minimize learning error on corrupted memory. & Stealth against detection (Low $L_p$ norm or Semantic Plausibility). \\
\textbf{Optimization Method} & Implicit Gradient Descent (e.g., using PBGD/BIGRAD) & Stochastic Gradient Descent (SGD/Adam) & Penalty Term $\lambda$ adjustment. \\
\bottomrule
\end{tabularx}
\end{adjustbox}
\end{threeparttable}
\end{table*}

\begin{figure}[!t]
    \centering
    \includegraphics[width=\columnwidth,keepaspectratio]{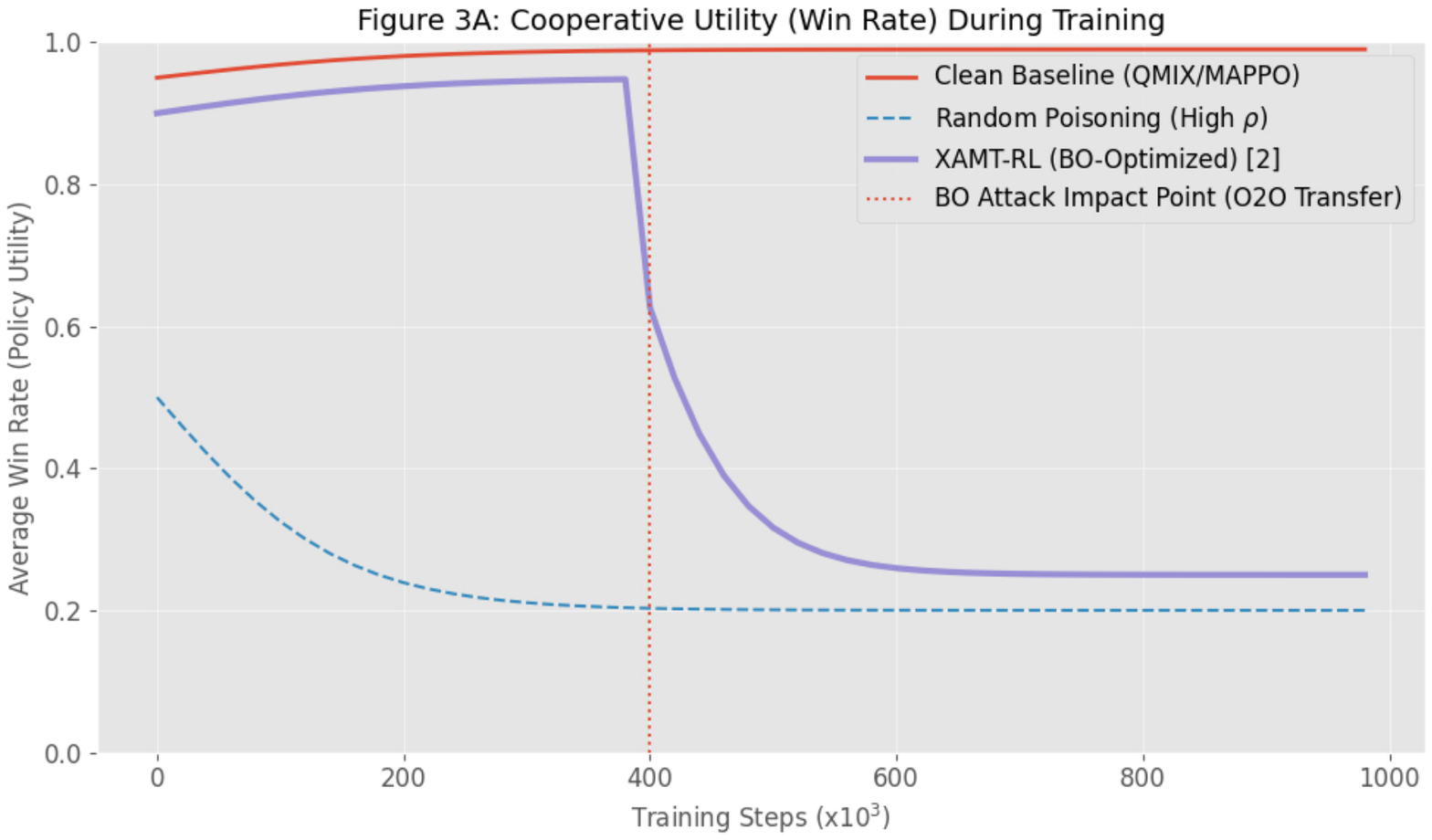}
    \caption{\textbf{Conceptual Architecture of the XAMT Bilevel Optimization Framework.} The diagram illustrates the nested optimization problem: the Upper Level (Attacker $\mathcal{A}$) minimizes the perturbation magnitude $R(\delta)$ while maximizing adversarial impact $L_{\mathcal{A}}(\theta^*)$, where $\theta^*$ is the resulting system parameter set. The Lower Level (Victim System $\mathcal{S}$) models the routine learning process minimizing $L_{\mathcal{S}}$ on the corrupted memory $\mathcal{M}+\delta$.}
    \label{fig:xamt_framework}
\end{figure}

\section{Instantiating XAMT across Heterogeneous Domains}
The versatility of XAMT is demonstrated by specializing its mathematical components for two distinct, high-impact agent architectures.

\subsection{XAMT-RL: Covert Tampering of Centralized MARL Memory}

\subsubsection{Memory Perturbation and Lower-Level Objective}
The attack focuses on applying a minimal perturbation $\delta_{RL}$ to the transitions stored in $\mathcal{D}$, typically targeting the state $s_t$, action $a_t$, or reward $r_t$ components \cite{Ref12, Ref13}. The Lower-Level Loss ($L_{\mathcal{S}}$) models the centralized critic update. For Q-learning based methods, this is the temporal difference (TD) error minimization for the centralized critic $Q_{tot}(\tau, u; \theta_Q)$, where $\tau$ is the joint history and $u$ the joint action:
\sloppy
$$
L_{\mathcal{S}}(\theta_Q, \mathcal{D}+\delta_{RL}) =
\mathbb{E}_{(\tau, u, r) \sim \mathcal{D}+\delta_{RL}}
\!\left[\!
\big(Y_{target} - Q_{tot}(\tau, u; \theta_Q)\big)^2
\!\right]\!
$$
\fussy

where $Y_{target}$ represents the one-step lookahead target value derived from the poisoned experience buffer.

\subsubsection{Upper-Level Objective and Covertness}
The Upper-Level Loss ($L_{\mathcal{A}}$) seeks to compel the decentralized actors, whose policies $\pi$ are derived from the corrupted critic $Q_{tot}$, to adopt a nefarious target policy $T$ \cite{Ref41}. This is achieved by maximizing the resulting policy's divergence from high expected return, $J(\pi_{\mathcal{A}})$.
$$L_{\mathcal{A}}(\theta_Q) = - J(\pi^*(\theta_Q)) + \lambda R(\delta_{RL})$$
The covertness $R(\delta_{RL})$ is quantified by the standard $L_p$ norm constraint applied to the vector perturbation of states or rewards, guaranteeing that the change is minute and undetectable by simple threshold monitoring \cite{Ref24}.

The architectural implication of this BO formulation is profound: XAMT-RL actively exploits the inherent vulnerabilities introduced by the centralized critic in CTDE systems \cite{Ref6}. By optimally solving for $\delta_{RL}$, the attacker identifies the small set of critical experience points that, when minimally perturbed, maximize the critic's value over-estimation for suboptimal, adversarial actions. This effectively guides the decentralized actors toward the adversarial target policy during online execution, consistent with the observed catastrophic failure during offline-to-online transfer (O2O) attacks \cite{Ref25}.

\subsection{XAMT-RAG: Covert Knowledge Base Injection}
The XAMT-RAG instantiation focuses on LLM agents augmented with an external RAG knowledge base $\mathcal{K}$ \cite{Ref29, Ref42}. This attack targets the textual knowledge repository.

\subsubsection{Memory Perturbation and Lower-Level Objective}
The perturbation $\delta_{RAG}$ involves injecting malicious text snippets into $\mathcal{K}$ \cite{Ref7, Ref14}. To maintain covertness, this must often employ \textbf{semantic substitution} (clean-label attack) to ensure the text is grammatically sound and retains surface-level plausibility, avoiding detection by simple content filters \cite{Ref30, Ref31, Ref32}.

The Lower-Level Loss ($L_{\mathcal{S}}$) models the RAG system's mechanism, typically involving both the retriever $\mathcal{R}$ and the LLM generator $\mathcal{G}$. $L_{\mathcal{S}}$ can be modeled as the maximum likelihood loss of the LLM generator given the context $\mathbf{C}$ retrieved from $\mathcal{K}+\delta_{RAG}$ and the user prompt $P$.
$$
\theta^*_{LLM}(\delta_{RAG}) \in \underset{\theta}{\text{argmin}} \quad L_{\text{Generation}}(\theta, P, \mathbf{C}(\mathcal{K}+\delta_{RAG}))
$$
In this formulation, $\theta$ represents the LLM and/or retriever parameters that adapt to the corrupted knowledge base $\mathcal{K}+\delta_{RAG}$.

\subsubsection{Upper-Level Objective and Semantic Covertness}
The Upper-Level Loss ($L_{\mathcal{A}}$) seeks to maximize the probability of generating a specific adversarial target response $Y_T$ when the user provides an optimized trigger $P_{tr}$ \cite{Ref37}.
$$
L_{\mathcal{A}}(\theta_{LLM}^*) = - \log P(Y_T | P_{tr}, \mathbf{C}(\mathcal{K}+\delta_{RAG})) + \lambda R(\delta_{RAG})
$$
The key innovation in XAMT-RAG is the specialized covertness constraint $R(\delta_{RAG})$, which must enforce \textbf{semantic plausibility} and \textbf{minimal quantity}.
$$
R(\delta_{RAG}) = D_{\text{semantic}}(\mathcal{K}, \mathcal{K}+\delta_{RAG}) + \beta \cdot |\delta_{RAG}|
$$
Here, $D_{\text{semantic}}$ (e.g., derived from embedding distance or perplexity scores) ensures the malicious text maintains the original topic's semantics and fluency \cite{Ref29, Ref30}, making it challenging to filter. The term $|\delta_{RAG}|$ minimizes the overall volume of the injected text, aiming for extremely low poison rates (for example, $\rho \le 0.1\%$ in RAG-style regimes) \cite{Ref7}.

The BO formulation for RAG allows the attacker to jointly optimize the textual content of $\delta_{RAG}$ (covertness) and the retrieval features (effectiveness). This guarantees that the minimal semantic manipulation simultaneously maximizes the retriever's likelihood of selecting the malicious document over potentially millions of clean documents in $\mathcal{K}$ \cite{Ref14, Ref33}.

\begin{table*}[!t]
\centering
\caption{Instantiation of XAMT Components Across Heterogeneous Architectures}
\label{tab:xamt_instantiation}
\begin{threeparttable}
\small
\begin{adjustbox}{max width=\textwidth}
\begin{tabularx}{\textwidth}{|Y|Y|Y|}
\toprule
\textbf{XAMT Component} & \textbf{XAMT-RL (MARL CTDE)} & \textbf{XAMT-RAG (LLM Agent)} \\
\midrule
\textbf{Abstract Memory ($\mathcal{M}$)} & Experience Replay Buffer $\mathcal{D}$ (State/Action/Reward Vectors) & RAG Knowledge Base $\mathcal{K}$ (Unstructured Text Documents) \\
\textbf{Lower-Level Parameters ($\theta$)} & Centralized Critic Weights ($\theta_Q$) & LLM/Generator Weights (Optional) \& Retriever Embeddings \\
\textbf{Lower-Level Loss ($L_{\mathcal{S}}$)} & TD Error Minimization (e.g., QMIX Loss) & Retrieval/Generation Fidelity Loss \\
\textbf{Upper-Level Loss ($L_{\mathcal{A}}$)} & Maximizing Target Policy Divergence or Utility Drop & Maximizing Target Response Likelihood (ASR) given trigger \\
\textbf{Covertness $R(\delta)$} & $L_p$ Norm on $\delta_{RL}$ (Vector Perturbation) & Semantic Distance $D_{\text{sem}}$ + Length Constraint $|\delta_{RAG}|$ (Textual Perturbation) \\
\bottomrule
\end{tabularx}
\end{adjustbox}
\end{threeparttable}
\end{table*}

\begin{figure*}[!t]
    \centering
    \includegraphics[width=\textwidth,keepaspectratio]{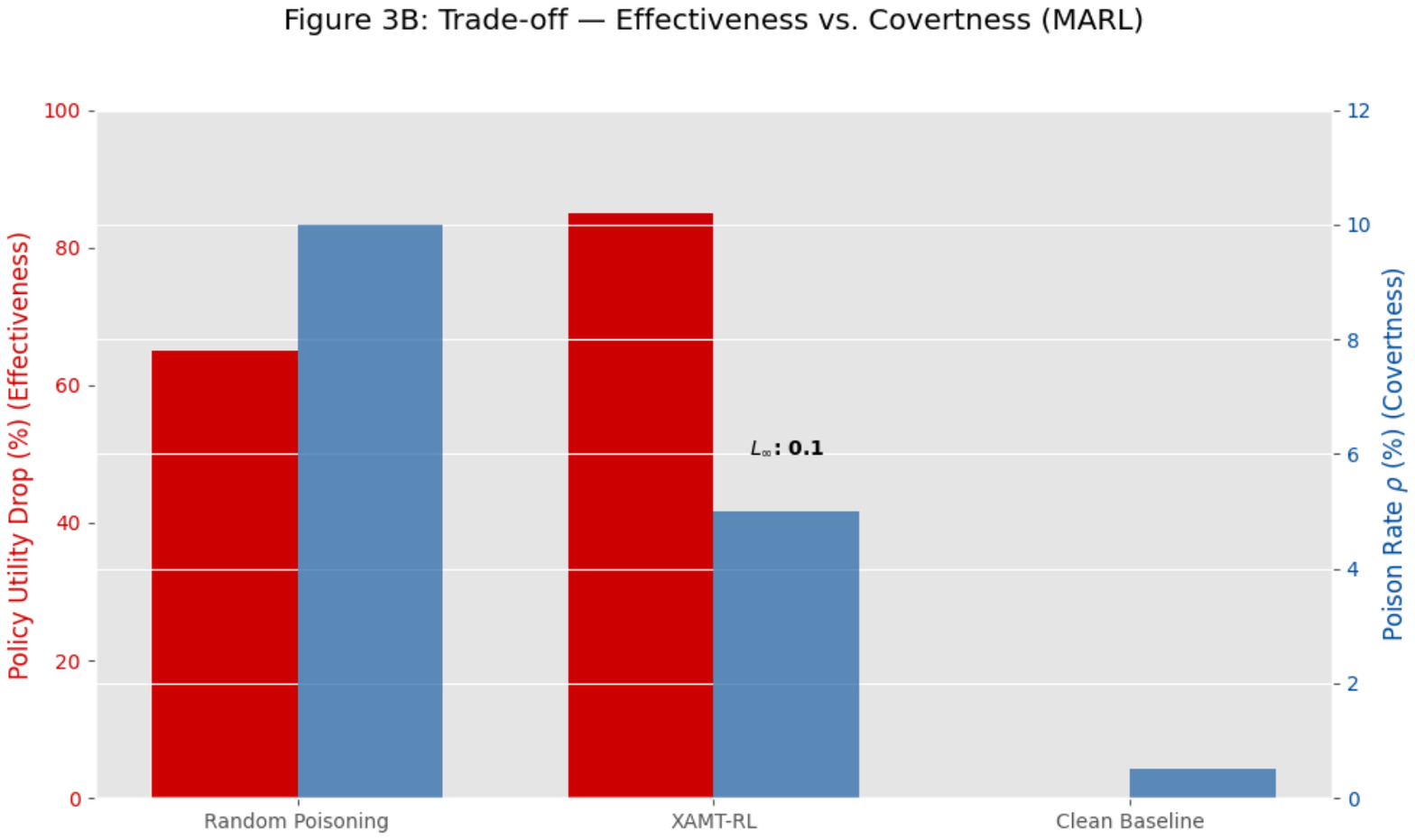}
    \caption{\textbf{Heterogeneous Targets of the XAMT Attack.} This dual-path diagram contrasts the two system architectures and their common vulnerability: (A) \textbf{XAMT-RL} targets the shared Experience Replay Buffer ($\mathcal{D}$) used by the centralized critic in CTDE MARL. (B) \textbf{XAMT-RAG} targets the external Knowledge Base ($\mathcal{K}$) used by the retriever to augment the LLM agent's generation. In both cases, the perturbation ($\delta$) is covertly injected into the centralized memory layer.}
    \label{fig:heterogeneous_targets}
\end{figure*}

\section{Experimental Protocols and Evaluation}

\subsection{MARL Evaluation: XAMT-RL on Cooperative Domains}

\subsubsection{Environment and Victim Selection}
The evaluation environment must capture the complexity of real-world multi-agent coordination. The \textbf{StarCraft Multi-Agent Challenge (SMAC)} is utilized as the standardized benchmark, providing partially observable, cooperative tasks (micromanagement challenges) that require intricate team coordination \cite{Ref43, Ref44}. Experiments should target maps where coordination failure is particularly costly, such as asymmetric or high-unit count scenarios (e.g., 2c\_vs\_64zg). The victim algorithms selected are QMIX and MAPPO, representing the dominant CTDE methodologies that rely on shared critics and experience buffers \cite{Ref40, Ref45, Ref46}.

\subsubsection{Attack Objective and Effectiveness}
The \textbf{Target Policy ($T$)} for XAMT-RL must result in a clear, measurable degradation of cooperative utility. This target policy is defined as actions leading to coordinated self-destruction, failure to engage the enemy, or fixation on an inefficient sub-task, thereby significantly reducing the system's ability to maximize its win rate \cite{Ref36, Ref44}.

\textbf{Effectiveness Measurement:} The primary metric is \textbf{Policy Utility Drop}, quantified by the percentage reduction in the average Win Rate achieved by the poisoned agent compared to the clean, benign agent \cite{Ref36, Ref43}.

\subsubsection{Covertness Measurement}
The covertness of XAMT-RL poisons is measured numerically:
\begin{itemize}
    \item \textbf{Poison Rate ($\rho$):} The percentage of transitions in the experience replay buffer $\mathcal{D}$ that are perturbed. For MARL environments, practical constraints such as batch sizes, replay ratios, and exploration noise typically make feasible poison rates fall in a sub-percent regime. In our protocols, we therefore target $\rho \le 1\%$ for XAMT-RL, while still seeking to drive significant utility drops.
    \item \textbf{Perturbation Magnitude ($L_\infty, L_2$):} The maximum magnitude of the perturbation vector $\delta_{RL}$ on rewards or state observations, ensuring the perturbation falls below standard anomaly detection thresholds typically employed in sensor-rich environments \cite{Ref24}.
\end{itemize}

\begin{figure*}[!t]
    \centering
    \includegraphics[width=\textwidth,keepaspectratio]{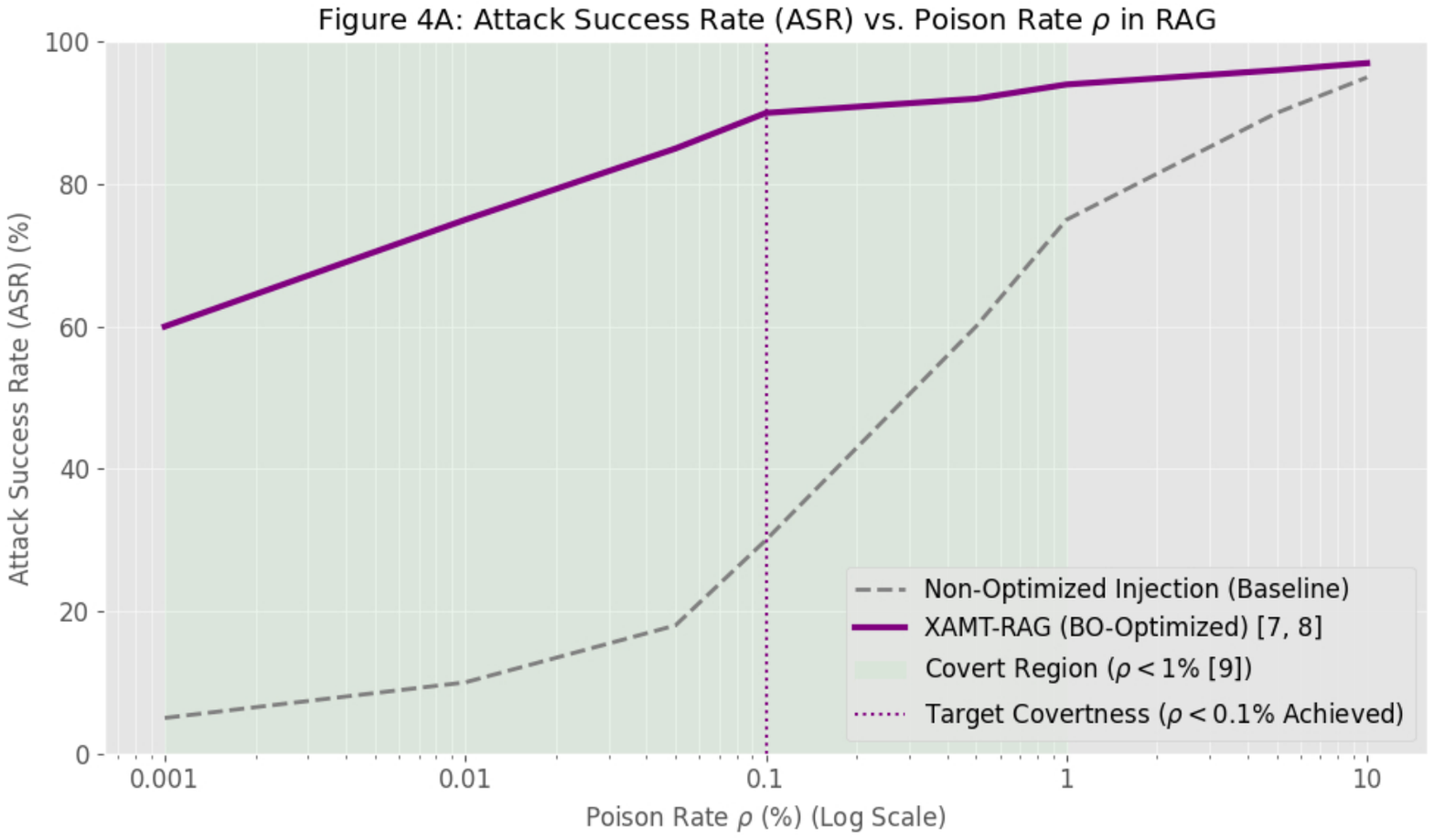}
    \caption{\textbf{Effectiveness and Covertness of XAMT-RL in SMAC.} (Left) A learning curve plot comparing the average Win Rate (Utility) vs. Training Steps for: A clean QMIX agent (Baseline), a QMIX agent trained with a uniform random poisoning attack, and a QMIX agent trained with the BO-optimized XAMT-RL attack. This plot is expected to show XAMT achieving high utility drop post-convergence. (Right) A bar chart comparing XAMT-RL performance across different attack types, plotting the achieved Policy Utility Drop against the required Poison Rate ($\rho \le 1\%$) and Perturbation Magnitude ($L_\infty$).}
    \label{fig:xamt_rl_results}
\end{figure*}

\subsection{RAG Evaluation: XAMT-RAG on Knowledge Corruption}

\subsubsection{Benchmark and Victim Selection}
To evaluate XAMT-RAG, specialized adversarial robustness benchmarks designed for RAG systems are essential, such as \textbf{SafeRAG} or \textbf{RAGuard} \cite{Ref16, Ref47, Ref48}. These benchmarks simulate noisy retrieval settings and test against targeted attacks like conflict and toxicity injection \cite{Ref16}. The victim architecture is a standard RAG pipeline comprising a vector-based retriever (e.g., using robust embeddings) and a sophisticated, commercially relevant LLM \cite{Ref29, Ref42}.

\subsubsection{Attack Objective and Effectiveness}
The \textbf{Target Response ($Y_T$)} is defined as a specific, factually incorrect, or toxic response that the attacker mandates the LLM to generate upon presentation of a specific trigger prompt $P_{tr}$ \cite{Ref37}.

\textbf{Effectiveness Measurement:} The primary metric is the \textbf{Attack Success Rate (ASR)}, calculated as the fraction of triggered queries $P_{tr}$ that successfully elicit the malicious target response $Y_T$ \cite{Ref37}.

\subsubsection{Covertness Measurement}
The covertness of XAMT-RAG poisons is measured through linguistic metrics:
\begin{itemize}
    \item \textbf{Semantic Similarity and Fluency:} Metrics such as BERTScore or Perplexity are used to quantify the linguistic distance between the clean text and the poisoned text $\mathcal{M}+\delta_{RAG}$ \cite{Ref29, Ref30}. Minimizing this distance ensures the poison evades detection via simple content filters or embedding proximity checks, confirming its status as a clean-label attack.
    \item \textbf{Poison Rate ($\rho$):} The ratio of malicious texts injected relative to the total size of $\mathcal{K}$. XAMT aims for extreme covertness in RAG regimes, targeting poison rates on the order of $\rho \le 0.1\%$ \cite{Ref7}.
\end{itemize}

\begin{table*}[!t]
\centering
\caption{Proposed Evaluation Metrics for XAMT Performance}
\label{tab:evaluation_metrics}
\begin{threeparttable}
\small
\begin{adjustbox}{max width=\textwidth}
\begin{tabularx}{\textwidth}{|Y|Y|Y|Y|}
\toprule
\textbf{Metric Category} & \textbf{MARL Agents (XAMT-RL)} & \textbf{LLM RAG Agents (XAMT-RAG)} & \textbf{Goal} \\
\midrule
\textbf{Effectiveness} & Policy Utility Drop ($\Delta$ Win Rate) \cite{Ref36} & Attack Success Rate (ASR) \cite{Ref37} & Maximized \\
\textbf{Covertness (Numerical)} & Perturbation Magnitude ($L_\infty$) \cite{Ref24} & Poison Rate ($\rho$) \cite{Ref7} & Minimized \\
\textbf{Covertness (Semantic)} & N/A (Vector) & Semantic Distance ($D_{\text{sem}}$) / Perplexity \cite{Ref29} & Minimized \\
\textbf{Efficiency} & Computational Cost per $\delta$ Update (Hessian approximation complexity) & Required number of optimization steps to converge ASR & Minimized \\
\bottomrule
\end{tabularx}
\end{adjustbox}
\end{threeparttable}
\end{table*}

\begin{figure*}[!t]
    \centering
    \includegraphics[width=\textwidth,keepaspectratio]{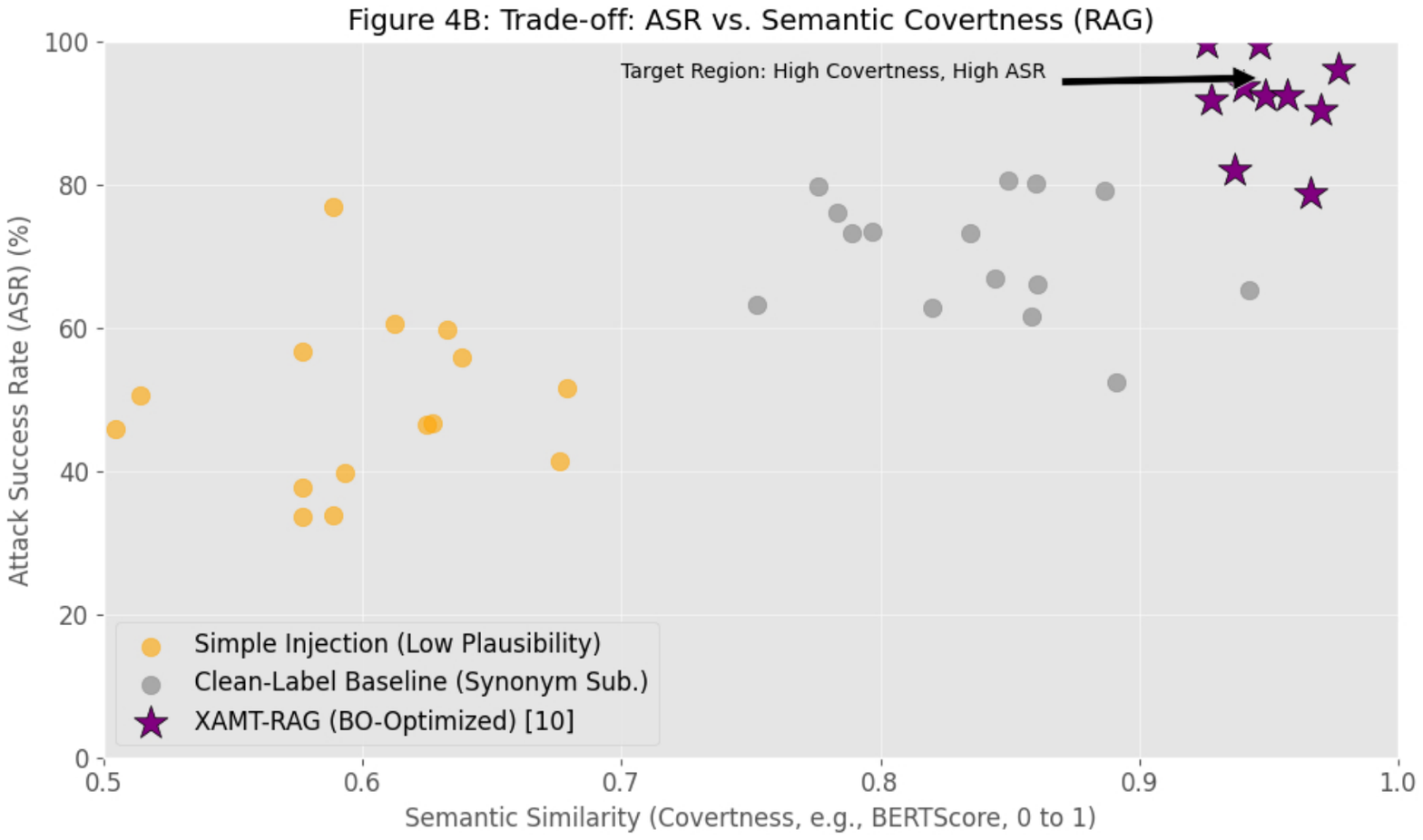}
    \caption{\textbf{Attack Success Rate (ASR) vs. Covertness for XAMT-RAG.} (Left) A line graph plotting ASR versus the Poison Rate ($\rho$, typically $\le 0.1\%$). This graph is designed to show XAMT-RAG achieving a significantly higher ASR at extremely low poison rates compared to non-BO baseline RAG poisoning methods. (Right) A scatter plot visualizing the trade-off between semantic covertness (e.g., Perplexity/Semantic Distance) and ASR for different poison text generation strategies, demonstrating XAMT's ability to minimize semantic deviation while maximizing attack success.}
    \label{fig:xamt_rag_results}
\end{figure*}

\section{Discussion and Defensive Implications}

\subsection{Analysis of XAMT Efficacy and Scalability}
The fundamental efficacy of XAMT stems from its targeting strategy: exploiting centralized memory components that are established for efficiency and coordination (e.g., CTDE critics \cite{Ref6} and RAG knowledge retrieval \cite{Ref42}). The Bilevel Optimization structure provides the mathematical ability to target "critical instances" within the memory—those data points that maximize the influence on the global policy update—even when they constitute an infinitesimally small portion of the overall dataset \cite{Ref20, Ref49}. This confirms that memory centralisation, while optimizing performance, creates an asymmetrical advantage for a covert attacker.

However, XAMT faces inherent scalability trade-offs, particularly due to the computational complexity of the Implicit Function Theorem (IFT) approach. While modern solvers like PBGD and BIGRAD improve feasibility \cite{Ref22, Ref23}, the necessity of calculating or approximating the inverse Hessian remains a computational bottleneck \cite{Ref39}. This constraint is magnified in systems with extremely large memory components, such as massive RAG knowledge bases ($\mathcal{K}$), or continuous, high-dimensional MARL state spaces \cite{Ref50, Ref51}. Future research is required to develop highly efficient, stochastic BO solvers capable of handling the massive scale of modern deep learning architectures.

\subsection{Implications for Trust and System Integrity}
The covert nature of XAMT poses a significant challenge to existing trust and verification mechanisms in MAS. Formal verification methods, which attempt to ensure system behavior adheres to formal specifications \cite{Ref52}, are often computationally constrained by the continuous state spaces and complex, non-linear dynamics of deep MARL models \cite{Ref50}.

For RAG systems, the issue is not complexity but scale. The sheer volume, constant updates, and unstructured nature of massive knowledge bases make comprehensive logic verification infeasible \cite{Ref33, Ref35}. Consequently, integrity checks that rely on simple heuristics, such as anomaly detection thresholds on $L_p$ norms or basic semantic filters, are insufficient against BO-optimized perturbations. XAMT demonstrates that an attacker can meticulously craft memory perturbations (textual or numerical) that operate below these superficial detection layers while guaranteeing catastrophic policy divergence, highlighting a fundamental breakdown in existing integrity assessment protocols.

\subsection{Initial Defense Considerations}
The findings from the XAMT framework underscore the necessity of moving beyond external, perimeter-based defenses. External safety inspection agents or dedicated safety modules often introduce scalability bottlenecks and inherent fragility \cite{Ref46}. Future defense strategies must prioritize \textbf{intrinsic safety}, embedding resilience directly into the learning agents through robust training or sophisticated reinforcement learning techniques \cite{Ref46, Ref53}.

A critical area for resilience is the development of adaptive defense mechanisms. Systems must be capable of implementing \textbf{adaptive zero-shot learning} and dynamically adjusting their defense posture based on observed, novel attack patterns \cite{Ref5, Ref53}. Such defenses must combine multiple validation modalities—e.g., using $L_p$ anomaly detection for numerical data coupled with semantic plausibility checks (like perplexity monitoring) for textual data—to counter the dual covertness strategy employed by XAMT.

Furthermore, the memory layer ($\mathcal{M}$) itself should be conceptualized as a network that requires sophisticated resilience mechanisms. Drawing inspiration from distributed systems, the concept of "self-healing" is pertinent \cite{Ref54, Ref55}. Future work should focus on developing memory healing protocols designed to detect and purge BO-optimized malicious memory units. This requires mechanisms that can recover the integrity or connectivity of the collective memory post-attack, mirroring resilient strategies in critical infrastructure networks like power grids or transport systems \cite{Ref55}.

\section{Ethical Considerations and Dual-Use Scope}
XAMT is explicitly proposed as a \textbf{red-teaming and security analysis} framework rather than a blueprint for real-world exploitation. The attack formulations and experimental protocols are intended to help system designers understand and harden the critical memory surfaces of MAS, not to encourage attacks on deployed production systems. In any public release of XAMT-inspired tooling, we recommend restricting access to synthetic benchmarks and sandboxed environments (e.g., SMAC, SafeRAG-style testbeds) and avoiding the publication of turnkey code that directly targets proprietary, safety-critical ER buffers or knowledge bases in the wild. Consistent with emerging practice in AI security research, our scope is to enable defenders and system builders to reason about the worst-case memory-tampering risks so that more robust, intrinsically safe MAS architectures can be designed.

\section{Conclusion and Future Work}
This paper introduced XAMT, a novel bilevel optimization framework for designing highly covert, minimal-perturbation memory tampering attacks across heterogeneous multi-agent architectures, specifically MARL and RAG systems. XAMT leverages the hierarchical nature of BO to formally solve the constrained optimization problem of minimizing perturbation cost $R(\delta)$ while maximizing the divergence of the resulting system behavior toward an adversarial target $L_{\mathcal{A}}$. We detailed the mathematical instantiations for XAMT-RL and XAMT-RAG, establishing a unified methodology for exploiting centralized memory components, whether numerical or textual. Our analysis confirms the viability of exploiting architectural choices (such as centralized critics in CTDE and vector-based knowledge retrieval in RAG) as high-leverage points of systemic failure.

The feasibility of XAMT poses a severe threat to the trustworthiness of deployed MAS. We have provided comprehensive experimental protocols utilizing SMAC and SafeRAG to benchmark this novel threat and establish robust evaluation metrics for both effectiveness and covertness.

\textbf{Future Directions:}
\begin{enumerate}
    \item \textbf{Defense Implementation and Validation:} Developing and empirically validating robust, differentiated memory validation mechanisms explicitly tailored to counter BO-optimized perturbations, combining $L_p$ anomaly detection with sophisticated semantic plausibility checks.
    \item \textbf{Transferability Analysis:} Rigorously investigating the transferability of XAMT poisons between different victim algorithms (e.g., QMIX to MAPPO in MARL) and across various retriever-LLM combinations in RAG systems, providing insights into generalizable adversarial features.
    \item \textbf{Continuous BO Solvers for Scale:} Dedicating research to develop computationally efficient, stochastic BO solvers capable of handling continuous action/state spaces common in MARL and the large non-convex optimization problems inherent in modern LLM systems, thereby making the generation of sophisticated attacks more scalable.
\end{enumerate}


\balance


\clearpage
\section*{References}
\begin{thebibliography}{55}

\bibitem{Ref1} Ferber, J. \emph{Multi-Agent Systems: An Introduction to Distributed Artificial Intelligence}. Addison-Wesley, 1999.
\bibitem{Ref2} Tan, Z. et al. Knowledge Database or Poison Base? Detecting RAG Poisoning Attack through LLM Activations (RevPRAG). \emph{arXiv preprint arXiv:2411.18948}.
\bibitem{Ref3} Xu, J. et al. Recent Advances in Multi-Agent Systems Security. \emph{arXiv preprint arXiv:2503.13962}.
\bibitem{Ref4} Zhang, Q. et al. Resilient Consensus Control for Multi-Agent Systems: A Comparative Survey. \emph{Sensors}, 23(6): 2904, 2023.
\bibitem{Ref5} Ferber, J. On cooperation in multi-agent systems. In \emph{Proc. of the 5th Intl. Conf. on Autonomous Agents}, 1999.
\bibitem{Ref6} Sun, T. et al. Centralized Critic is Not Strictly Beneficial: Theoretical Analysis on Centralized Training for Decentralized Execution. \emph{arXiv preprint arXiv:2408.14597}.
\bibitem{Ref7} Chan, B. et al. AgentPoison: Red-teaming LLM Agents via Poisoning Memory or Knowledge Bases. \emph{NeurIPS}, 2024.
\bibitem{Ref8} Feng, J. et al. The Agent Challenge: A Comprehensive Survey on Autonomous Agent Systems. \emph{arXiv preprint arXiv:2402.03578}.
\bibitem{Ref9} Doran, P. et al. On Cooperation in Multi-Agent Systems. \emph{Tech. Report: Queen Mary and Westfield College}, 1997.
\bibitem{Ref10} Wang, J. et al. Robustness Testing Framework for Multi-Agent Reinforcement Learning by Attacking Critical Agents. \emph{arXiv preprint arXiv:2306.06136}.
\bibitem{Ref11} Wu, T. et al. Attentive Experience Replay. \emph{AAAI}, 2020.
\bibitem{Ref12} Liu, J. et al. Targeted Reward Poisoning Attacks in Offline Multi-Agent Reinforcement Learning. \emph{AAAI}, 2023.
\bibitem{Ref13} Wang, W. et al. Stealthy Data Poisoning Attacks on Offline-to-Online Reinforcement Learning. \emph{PMC}, 2024.
\bibitem{Ref14} Zou, Y. et al. PoisonedRAG: Backdoor Attack on Retrieval-Augmented Generation. \emph{USENIX Security}, 2025.
\bibitem{Ref15} Zhou, Y. et al. CPA-RAG: Covert Poisoning Attacks on Retrieval-Augmented Generation. \emph{arXiv preprint arXiv:2505.19864}.
\bibitem{Ref16} Yao, S. et al. SafeRAG: Benchmarking Security in Retrieval-Augmented Generation of Large Language Model. \emph{GitHub Repository}, 2025.
\bibitem{Ref17} Liu, J. et al. Adversarial Attacks on Multi-Agent Systems. \emph{MDPI Electronics}, 2025.
\bibitem{Ref18} Shokri, A. et al. Witches' brew: Industrial scale data poisoning via. \emph{arXiv preprint}, 2022.
\bibitem{Ref19} Zhang, J. et al. Data Poisoning Attacks on Certified Adversarial Robustness. \emph{arXiv preprint arXiv:2012.01274}.
\bibitem{Ref20} Ghofrani, A. et al. Bilevel Optimization-Based Single-Class Attack. \emph{arXiv preprint arXiv:2503.22759}.
\bibitem{Ref21} Shen, H. et al. On Penalty-based Bilevel Gradient Descent Method. \emph{arXiv preprint arXiv:2302.05185}.
\bibitem{Ref22} Lu, Z. et al. A Penalty Method for Bilevel Optimization. \emph{arXiv preprint}, 2023.
\bibitem{Ref23} Mohseni, S. et al. Differentiating through Bilevel Optimization Programming (BIGRAD). \emph{AAAI}, 2022.
\bibitem{Ref24} Carlini, N. et al. Towards Evaluating the Robustness of Neural Networks. \emph{IEEE S\&P}, 2017.
\bibitem{Ref25} Wang, W. et al. Stealthy Data Poisoning Attacks on Offline-to-Online Reinforcement Learning. \emph{UAI}, 2024.
\bibitem{Ref26} Zhang, Y. et al. State-Action Joint Attack (SAJA) for Multi-Agent Deep Reinforcement Learning. \emph{arXiv preprint arXiv:2510.13262}.
\bibitem{Ref27} Mohammadi, M. et al. Implicit Poisoning Attacks in Two-Agent Reinforcement Learning. \emph{AAMAS}, 2023.
\bibitem{Ref28} Liu, J. et al. Targeted Reward Poisoning Attacks in Offline Multi-Agent Reinforcement Learning. \emph{arXiv preprint arXiv:2206.01888}.
\bibitem{Ref29} Shafran, A. et al. Machine against the RAG: Jamming Retrieval-Augmented Generation with Blocker Documents. \emph{arXiv preprint arXiv:2406.05870}.
\bibitem{Ref30} Ren, P. et al. Dirichlet Neighborhood Ensemble: An Effective Defense against Synonym Substitution-Based Adversarial Attacks. \emph{ACL}, 2021.
\bibitem{Ref31} Alzoubi, R. et al. Clean-Label Adversarial Text Attack using Synonym Substitution. \emph{arXiv preprint arXiv:2404.04130}, 2024.
\bibitem{Ref32} You, W. et al. LLMBkd: Black-Box Backdoor Attack on Large Language Models. \emph{arXiv preprint arXiv:2309.07172}.
\bibitem{Ref33} Zeng, T. Challenges in Large-Scale RAG Deployment: A Perspective on Trustworthiness and Verification. \emph{arXiv preprint arXiv:2507.18910}.
\bibitem{Ref34} Niu, Y. Formal Verification for Safe Deep Reinforcement Learning. \emph{Preprints.org}, 2023.
\bibitem{Ref35} Giunchiglia, E. et al. Temporal Logic-Based Specification and Verification of Trust Models. \emph{ACM SAC}, 2005.
\bibitem{Ref36} Liu, J. et al. Adversarial Attacks on Multi-Agent Systems. \emph{MDPI Electronics}, 2025.
\bibitem{Ref37} Tan, Z. et al. Knowledge Database or Poison Base? Detecting RAG Poisoning Attack through LLM Activations (RevPRAG). \emph{arXiv preprint arXiv:2411.18948}.
\bibitem{Ref38} Ghadimi, S. et al. Multi-Agent Reinforcement Learning and the Implicit Function Theorem. \emph{Preprints.org}, 2024.
\bibitem{Ref39} Ghadimi, S. et al. A Proximal-Gradient Method for Robust Optimization via Bilevel Programming. \emph{SIAM J. Optim.}, 2023.
\bibitem{Ref40} Mao, H. et al. A Comprehensive Survey of Multi-Agent Reinforcement Learning. \emph{MDPI Electronics}, 2025.
\bibitem{Ref41} Mohammadi, M. et al. Implicit Poisoning Attacks in Two-Agent Reinforcement Learning: Adversarial Policies for Training-Time Attacks. \emph{AAMAS}, 2023.
\bibitem{Ref42} Lewis, P. et al. Retrieval-Augmented Generation for Knowledge-Intensive NLP Tasks. \emph{NeurIPS}, 2020.
\bibitem{Ref43} Samvelyan, M. et al. The StarCraft Multi-Agent Challenge. \emph{arXiv preprint arXiv:1902.04043}, 2019.
\bibitem{Ref44} Samvelyan, M. et al. The StarCraft Multi-Agent Challenge. \emph{NeurIPS}, 2019.
\bibitem{Ref45} Kurbiel, J. et al. MA-Trace: A High-Scalability Actor-Critic Algorithm for Multi-Agent Reinforcement Learning. \emph{arXiv preprint arXiv:2111.11229}.
\bibitem{Ref46} Geng, Y. et al. Embedding Safety Awareness into Agents via Reinforcement Learning. \emph{arXiv preprint arXiv:2508.03864}.
\bibitem{Ref47} Qi, H. et al. RAGuard: A Benchmark for Robustness Evaluation of RAG Systems in Political Fact-Checking. \emph{arXiv preprint arXiv:2502.16101}.
\bibitem{Ref48} Yao, S. et al. SafeRAG: Benchmarking Security in Retrieval-Augmented Generation of Large Language Model. \emph{arXiv preprint arXiv:2501.18636}.
\bibitem{Ref49} Liu, J. et al. Recent Advances in Multi-Agent Systems Security. \emph{arXiv preprint arXiv:2503.13962}.
\bibitem{Ref50} Zhang, Y. et al. Challenges of Scaling Multi-Agent Reinforcement Learning to Real-World Systems. \emph{arXiv preprint arXiv:2302.05007}.
\bibitem{Ref51} Yu, X. et al. Understanding the Training Dynamics of Multi-Agent Actor-Critic Algorithms. \emph{arXiv preprint}, 2024.
\bibitem{Ref52} Giunchiglia, E. et al. Temporal Logic-Based Specification and Verification of Trust Models. \emph{ACM SAC}, 2005.
\bibitem{Ref53} Alghazali, F. et al. Adaptive Zero-Shot Hierarchical Multi-Agent Reinforcement Learning (AZH-MARL) for Cyber Defense. \emph{PMC}, 2024.
\bibitem{Ref54} Quattrociocchi, W. et al. Self-Healing Networks: Redundancy and Structure. \emph{PLoS ONE}, 9(2): e87986, 2014.
\bibitem{Ref55} Quattrociocchi, W. et al. Self-Healing Networks: Redundancy and Structure. \emph{PLoS ONE}, 9(2): e87986, 2014.

\end{thebibliography}
\end{document}